\documentclass[%
 reprint,
superscriptaddress,
nofootinbib,
 amsmath,amssymb,
onecolumn
]{revtex4-2}

\usepackage{amssymb}
\usepackage{amsmath}
\usepackage{amsthm}
\usepackage{mathrsfs}
\usepackage{graphicx}
\usepackage{dcolumn}
\usepackage{bm}
\usepackage{hyperref}
\usepackage[usenames,dvipsnames,x11names,table]{xcolor}
\usepackage{balance}
\usepackage{dsfont}
\usepackage{tikz}
\usetikzlibrary{shapes}
\usepackage{multirow}
\usepackage{tabularx}
\usepackage{enumitem}
\usepackage{cancel,soul,ulem}
\setlist[itemize]{align=parleft,left=0pt..1em}

\begin{document}


\title{\textbf{Analysis of the heat transfer fluctuations in the Rayleigh-Bénard convection of concentrated emulsions with finite-size droplets}}

\author{Francesca Pelusi}
\email{f.pelusi@iac.cnr.it}
\affiliation{Istituto per le Applicazioni del Calcolo, CNR - Via dei Taurini 19, 00185 Rome, Italy}
\author{Stefano Ascione}
\affiliation{Department of Physics, Tor Vergata University of Rome - Via della Ricerca Scientifica 1, 00133 Rome, Italy}
\author{Mauro Sbragaglia}
\affiliation{Department of Physics \& INFN, Tor Vergata University of Rome, Via della Ricerca Scientifica 1, 00133 Rome, Italy}
\author{Massimo Bernaschi}
\affiliation{Istituto per le Applicazioni del Calcolo, CNR - Via dei Taurini 19, 00185 Rome, Italy}


\vspace{0.5cm}
\date{\today}

\begin{abstract}
Employing numerical simulations, we provide an accurate insight into the of heat transfer mechanisms in the Rayleigh-B\'{e}nard convection of concentrated emulsions with finite-size droplets. We focus on the unsteady dynamics characterizing the thermal convection of these complex fluids close to the transition from conductive to convective states, where the heat transfer phenomenon, expressed in terms of the Nusselt number Nu, is characterized by pronounced fluctuations triggered by collective droplets motion [Pelusi {\it et al.}, {\it Soft Matter} {\bf 17}(13), 3709 - 3721 (2021)]. By systematically increasing the droplet concentration, we show how these fluctuations emerge along with the segregation of ``extreme events" in the boundary layers, causing intermittent bursts in the heat flux fluctuations. Furthermore, we quantify the extension $S$ and the duration $\mathcal{T}$ of the coherent droplet motion accompanying these extreme events via a suitable statistical analysis involving the droplets displacements. We show how the increase in droplet concentration results in a power-law behaviour of the probability distribution function of $S$ and $\mathcal{T}$ and how this outcome is robust at changing the analysis protocol. Our work offers a comprehensive picture, linking macroscopic heat transfer fluctuations with the statistics of droplets at the mesoscale.
\end{abstract}

\maketitle


\section{\label{sec:intro}Introduction}

Rayleigh-Bénard Convection (RBC) is one of the most paradigmatic buoyancy-driven flows in fluid dynamics~\cite{Benard1900,Rayleigh1916}. It is observed whenever a fluid is placed between two plates under the influence of buoyancy forces, while heated from below and cooled from above. The heat transfer properties are generally quantified in terms of the dimensionless Nusselt number Nu, expressing the importance of the convective transport in comparison to the conductive one~\cite{Ahlers09,Lohse10,Chilla12,Shishkina21}. RBC plays an important role in a variety of fields, ranging from the atmosphere dynamics~\cite{Stevens05,Shipley22,Lin22} to the design of indoor environments~\cite{Enescu17,Xu21}, from the geophysical context~\cite{LeBarsDavaille04} to the metallurgic industry~\cite{Matson22}. From the theoretical side, RBC represents a fascinating problem that leads to the study of the instabilities and the transition from conductive to convective states, with the associated heat transfer properties, from the large scales down to the small ones. Different reviews have been written on the topic, covering experimental, numerical, and theoretical aspects~\cite{Ahlers09,Lohse10,Chilla12,Plumley19}.\\
\begin{figure*}[t!]
    \centering
    \includegraphics[width=.92\linewidth]{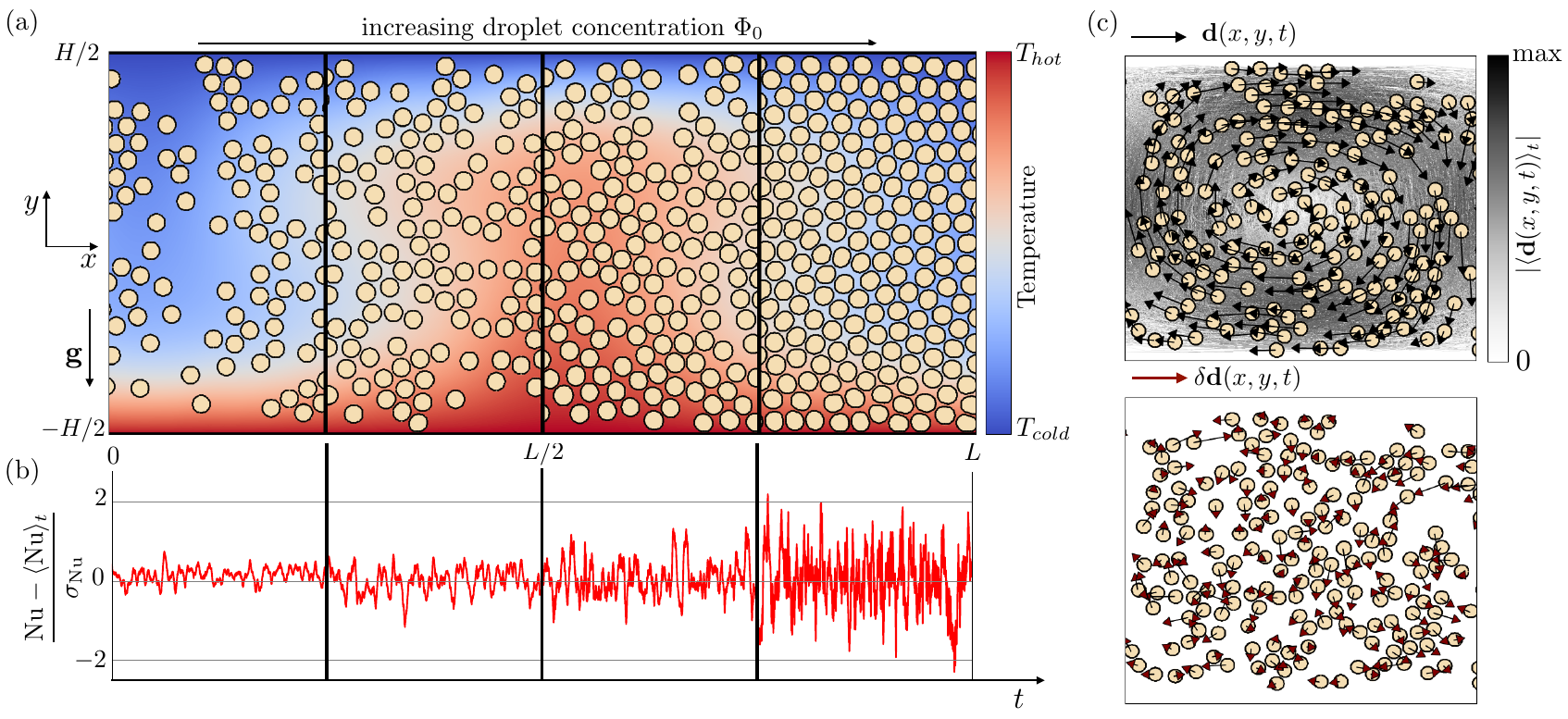}
    \caption{Panel (a): A sketch of the numerical setup. We investigate heat transfer fluctuations in a Rayleigh-B\'{e}nard cell with horizontal size $L$, where an emulsion is placed between a bottom hot and a top cold wall at a distance equal to $H$. In this setup, the emulsion dynamics is affected by the action of the gravity {$\bm g$} and by the presence of a temperature field, which varies from $T_{hot}$ (red) to $T_{cold}$ (blue), creating a temperature range $\Delta T=T_{hot}-T_{cold}$ between the two walls. The numerical experiment is repeated at increasing droplet concentration $\Phi_0$, from a less concentrated ($\Phi_0 = 0.27$) to a more concentrated emulsion ($\Phi_0=0.73$). Panel (b): we study the heat transfer in terms of the dimensionless parameter Nu defined in Eq.~\eqref{eq:NuTime}. Specifically, we are interested in fluctuations of Nu with respect to the averaged-in-time value $\langle \mathrm{Nu} \rangle_t$ (here shown normalised to its standard deviation $\sigma_\mathrm{Nu}$), which are observed to increase at increasing $\Phi_0$. Panel (c): A sketch of the computation of the Eulerian displacement fluctuations $\delta {\bf d}(x,y,t)$ defined in Eq.~\eqref{eq:displacementFluct} (bottom panel, brown arrows) as the difference between the Eulerian droplet displacement ${\bf d}(x,y,t)$ (top panel, black arrows) and the absolute value of the averaged-in-time displacement field $\langle {\bf d}(x,y,t) \rangle_t$ (top panel, colorbar).\label{fig:sketch}}
\end{figure*}
RBC has been traditionally addressed in the context of single-phase fluids, but studies in recent years also investigated the importance of the multi-phase and/or multi-component nature of the convective fluids~\cite{Wang07}, since it can impact engineering technological applications, such as energy storage~\cite{Moradikazerouni22,Du23,zhang2023}, petroleum industry~\cite{Ghorayeb2000,Gupta15}, and liquid food processing~\cite{Welti05,Moraga2011}. For example, RBC between two fluid layers has been studied, and numerical simulations helped in understanding the relationship between the heat transport efficiency and the properties of the two fluid layers, e.g., viscosity contrast, density contrast, and layers thickness~\cite{Yoshida16,Liu21,Liu22}. In the presence of multi-phase fluids, some studies also revealed enhanced heat transfer in comparison to the single-phase case, especially in proximity of the critical point, due to an increased occurrence of droplet condensation, or in the presence of the melting of a solid above a liquid melt~\cite{Zhong09,Yang23}. RBC laden with bubbles/droplets has also been numerically studied in recent works~\cite{Orestaetal09,Lohse10,Lakkaraju13,Biferale12,Li21,Liu21a,PelusiSM21}, showing how the heat transport properties can be affected by the presence of the dispersed phase~\cite{Orestaetal09,Lohse10,Biferale12,Lakkaraju13}, the surface wettability~\cite{Liu21a}, the condensation condition~\cite{Li21}, and the presence of non-trivial correlations between distant droplets~\cite{PelusiSM21}. Furthermore, experiments show that the introduction of a small percentage of a second component in a pure water solution is sufficient to affect the overall heat transfer~\cite{Wang19,Wang20}.\\
In the present paper, we focus on those situations where RBC laden with droplets is studied at increasing droplet concentration --  i.e., concentrated emulsions -- resulting in a non-Newtonian response of the fluid~\cite{Bonn17}. The impact of non-Newtonian rheology itself has been studied in the framework of RBC for a while~\cite{Zhang06,BalmforthRust09,Vikhansky09,Vikhansky10,AlbaalbakiKhayat11,Turanetal12,Davailleetal13,Massmeyer13,Kebicheetal14,Balmforthetal14,Hassanetal15,Karimfazli16}, while in a recent study~\cite{PelusiSM21} some of the authors have highlighted the richness brought by the finite-size of the droplets, i.e., the situation where RBC takes place in a confined environment and the actual extension of the droplets cannot be neglected in comparison to the characteristic size of the confining channel (cfr. Fig.~\ref{fig:sketch}(a)). Numerical simulations allowed to tune the droplet concentration, ranging from less concentrated Newtonian to more concentrated non-Newtonian emulsions, revealing that the increase in droplet concentration results in enhanced fluctuations in the heat transfer~\cite{PelusiSM21}. Although these fluctuations may exhibit qualitative similarities with fluctuations triggered in homogeneous fluids under turbulent
RBC~\cite{Aumaitre03,Labarre23}, they have the peculiarity of emerging just above the transition from a conductive to a convection state, i.e., in a purely laminar regime. This evidence highlights the role played by finite-size droplets. Here we take a step forward and provide a comprehensive characterization of the fluctuations. We first delve deeper into the connection between fluctuations at large scales and fluctuations at the droplet scales by systematically increasing the droplet concentration. The increase in droplet concentration reveals the emergence of ``extreme'' fluctuations that we characterize in terms of their spatial localization. The extreme fluctuations materialize in terms of collective droplet motion with a characteristic extension $S$ and duration $\mathcal{T}$. We therefore investigate the statistical properties of both $S$ and $\mathcal{T}$ following a well-established protocol based on the tracking of droplets displacement~\cite{PelusiAvalanche19}. This result is robust at changing analysis protocol.\\
The rest of the paper is organized as follows: the numerical methodology along with the necessary theoretical tools of analysis are recalled in Section~\ref{sec:methods}; a detailed characterization of the heat flux fluctuations, from ``extreme events" localization to their statistical characterization, is provided in Section~\ref{sec:analysis}; conclusions will be drawn in Section~\ref{sec:conclusions}.

\section{Methods}\label{sec:methods}
Numerical simulations have been conducted with the open-source code TLBfind~\cite{TLBfind22} developed by some of the authors. TLBfind is a GPU code that implements a lattice Boltzmann model~\cite{Kruger17,Succi18} for non-ideal multi-component systems to simulate the thermal convection of a multi-component fluid made of immiscible non-coalescing droplets in a two-dimensional system. The multi-component fluid comprises two components with densities $\rho_1$ and $\rho_2$, respectively~\footnote{Note that hereafter we use the convention of considering the first component as the one associated with the dispersed phase.}. Non-ideal interactions between the two components allow for phase separation and the formation of diffuse interfaces between regions with the majority of one of the two components, while additional competing interactions are considered to inhibit droplet coalescence. In TLBfind, the dynamics of the temperature field $T(x,y,t)$ obeys the advection-diffusion equation~\cite{Chandrasekhar61}, driven by the hydrodynamic fluid velocity. In turn, the two fluid components evolve following the Navier-Stokes equation with an additional buoyancy term introduced via the Boussinesq approximation~\cite{Spiegel60} implying a force term ${\bm F} =\rho \alpha {\bm g} T$, where $\rho=\rho_1+\rho_2$ is the total density, $\alpha$ the thermal expansion coefficient and $g$ the gravity acceleration. Furthermore, TLBfind allows the choice of the system size $L$ and $H$ along the $x$- and $y$- direction, respectively, and the number of emulsion droplets $N_{droplets}$ (cfr. Fig.~\ref{fig:sketch}(a)). The resulting droplet concentration $\Phi_0$ is defined as the fraction of domain size occupied by the dispersed phase, i.e., $\Phi_0 = \left\{\int \int \Theta(\rho_1(x,y)-\rho^*)\,dx \ dy \right\}/L H$, where $\Theta$ is the Heaviside step function and $\rho^*$ is a reference density value (cfr. Ref.~\cite{PelusiSM21}).  Relevant equations on the employed thermal multi-component lattice Boltzmann model are provided in Section~1 of the Supplementary Material.\\
In this work, simulations have been performed on a domain of size $H \sim 40 \ d$ and $L \sim 2H$, where $d$ is the average droplet diameter and it is fixed to 40 lattice spacing ($\Delta x$) for each simulated emulsion. Our choice of domain size allowed us to maintain consistency between simulations at different droplet concentrations. We explore emulsions with different $\Phi_0$ by varying $N_{droplets}$, i.e., from less concentrated ($\Phi_0 = 0.27$, $N_{droplets} = 352$) to more concentrated emulsions ($\Phi_0 = 0.73$, $N_{droplets} = 800$).
A detailed rheological characterization of the simulated emulsions is provided in Section~2 of the Supplementary Material. In all simulations, density and viscosity ratios between dispersed and continuous phases are fixed to the unity. Furthermore, the Capillary number (Ca) and Reynolds number (Re) are measured to be moderately small (Ca, $< 10^{-2}$, Re $< 10^{2}$).\\
We prepare monodisperse emulsions with a desired droplet concentration and we place them between two parallel plates located at $y=\pm H/2$. In the proximity of these plates, both components feel the effect of no-slip velocity boundary conditions, while we linearly initialize the temperature field between the two walls by prescribing $T(x,y=\pm H/2,t)=\mp \Delta T/2$, resulting in a system heated from below and cooled from above~\footnote{The value of $\Delta T = T_{hot} - T_{cold}$ (cfr. Fig.~\ref{fig:sketch}(a)) is fixed in all simulations.}. Notice that, before applying the buoyancy term to the two fluid components dynamics, a ``preparation'' run is necessary to let the emulsion relaxing towards its equilibrium configuration.
Intending to investigate the heat transfer properties of the emulsions in the framework described above, we introduce a macroscopic observable indicating the importance of convective transport, i.e., the dimensionless Nusselt number. This quantity is defined as~\cite{Shraiman90,Ahlers09,Verzicco10,Chilla12}
\begin{equation}\label{eq:NuTime}
\mathrm{Nu}(t) = \frac{\langle u_y (x,y,t) T (x,y,t)\rangle_{x,y} - \kappa \langle \partial_y T(x,y,t) \rangle_{x,y}}{\kappa \frac{\Delta T}{H}} \ ,
\end{equation}
where $u_y$ is the $y$-component of the hydrodynamical velocity field, $\kappa$ is the thermal diffusivity (see Section~1 of the Supplementary Material), and the angular brackets indicate a spatial average of the total domain size. We remark that a value of Nu equal to unity implies a conductive state, whereas a value larger than unity implies convective transport. For a homogeneous (single-phase) fluid, it is well established that the destabilization of a conductive state results in a convective state characterized by both a steady flow and a value of Nu that is time-independent. Contrariwise, for heterogeneous fluids, particularly in the case of concentrated emulsions, $\mathrm{Nu}$ exhibits fluctuations in time around a time-averaged value $\langle\mathrm{Nu}\rangle_t$~\cite{PelusiSM21} (see Fig.~\ref{fig:sketch}(b)), due to the presence of finite-size droplets. To characterize these fluctuations, for each emulsion, we have carefully chosen the amplitude of the buoyancy force to keep the time-averaged value $\langle\mathrm{Nu}\rangle_t \approx 2$, i.e., just above the transition from conduction to convection. This calibration of the buoyancy force to keep $\langle\mathrm{Nu}\rangle_t$ unchanged is required because, by increasing the concentration, a shear-thinning rheological behaviour of the emulsion emerges and the viscosity acquires a dependence on the shear rate. Overall, the ``effective'' viscosity $\eta_{eff}$ increases, and the ``effective'' Rayleigh number $Ra_{eff} = \rho \alpha g \Delta T H^3/(\kappa \eta_{eff})$ decreases unless $\alpha g$ is increased. As remarked in Ref.~\cite{PelusiSM21}, providing a precise estimation of the effective viscosity is a non trivial task due to the lack of a precise protocol. We may in general refer to a nominal Rayleigh number computed considering the dynamic viscosity $\eta_0$ of the continuous phase (see Section~1 of the Supplementary Material), i.e., $Ra = \rho \alpha g \Delta T H^3/(\kappa \eta_0)$, that varies between $6.7 \ 10^{3}$ (less concentrated emulsion) and $8 \ 10^{4}$ (more concentrated emulsion). In this laminar regime, droplet coalescence is not observed. The comparison between emulsion systems whose heat transfer is, on average, the same has the aim to highlight how a systematic variation in the emulsion concentration influences the fluctuations of Nu. This analysis will be conducted at scales comparable to the size of the droplets. Specifically, we introduce the single droplet Nusselt number $\mathrm{Nu}_{i}(t)$, i.e., a value of Nu associated with the $i$-th droplet, which can be obtained following Ref.~\cite{PelusiSM21} as a result of the decomposition of Eq.~\eqref{eq:NuTime} into the contributions of each single droplet
\begin{equation}\label{eq:NuDrops}
\mathrm{Nu}_{i}^{(drop)}(t) = \frac{u^{(i)}_y(t)  T^{(i)}(t) - \kappa (\partial_y T)^{(i)}(t)}{\kappa \frac{\Delta T}{H}}.
\end{equation}
For the sake of simplicity, hereafter we will refer to $\mathrm{Nu}^{*,(drop)}_{i}$ as the fluctuations of $\mathrm{Nu}_{i}^{(drop)}$ with respect to its averaged value $\langle \mathrm{Nu}^{(drop)}\rangle$ and normalized with respect to its standard deviation $\sigma_{\mathrm{Nu}}$:
\begin{equation}\label{eq:nu_drop_star}
    \mathrm{Nu}_i^{*,(drop)} (t) = \dfrac{\mathrm{Nu}_i^{(drop)}(t)- \langle \mathrm{Nu}^{(drop)}\rangle}{\sigma_{\mathrm{Nu}}}.
\end{equation}
$\langle \mathrm{Nu}^{(drop)}\rangle$ as well as $\sigma_{\mathrm{Nu}}$ result from considering all droplets $i$ at all times. Besides the droplet heat flux fluctuations in Eq.~\eqref{eq:nu_drop_star},
another core observable for the purpose of this work is the droplet displacement: by exploiting the Lagrangian tool embedded in TLBfind, which can individually track all the droplets via the identification of their centers of mass ${\bm X}_i= X_i\hat{x} + Y_i\hat{y}$, we compute the vectorial displacement ${\bf d}_i(t)$. Then, starting from ${\bf d}_i(t)$, at any simulation time step, the corresponding Eulerian field ${\bf d}(x,y,t)$ is extracted~\cite{PelusiSM21}. Finally, we compute the averaged-in-time displacement field $\langle {\bf d}(x,y,t) \rangle_t$ and consider the fluctuations of ${\bf d}(x,y,t)$ with respect to it
\begin{equation}\label{eq:displacementFluct}
\delta {\bf d}(x,y,t)={\bf d}(x,y,t)-\langle {\bf d}(x,y,t) \rangle_t .
\end{equation}
Fig.~\ref{fig:sketch}(c) shows a sketch of the above-mentioned fields. Note that the displacement fluctuation in Eq.~\eqref{eq:displacementFluct} is computed in a time range that is large enough to collect sufficient statistics but limited to an interval where the thermal plume does not move too much in the $x$ direction. \\

All the simulations have been conducted on GPUs (Tesla K80 and Quadro RTX 8000 GPUs), gathering data about tens of millions of droplets for each emulsion. In order to collect the same amount of data for all the concentrations, less concentrated emulsions require longer simulations due to the smaller number of droplets for each time step. In all cases, we exclude from the statistical analysis the initial, transient, period necessary to the development of the convective state. 
\begin{figure*}[t!]
    \centering
    \includegraphics[width=.81\linewidth]{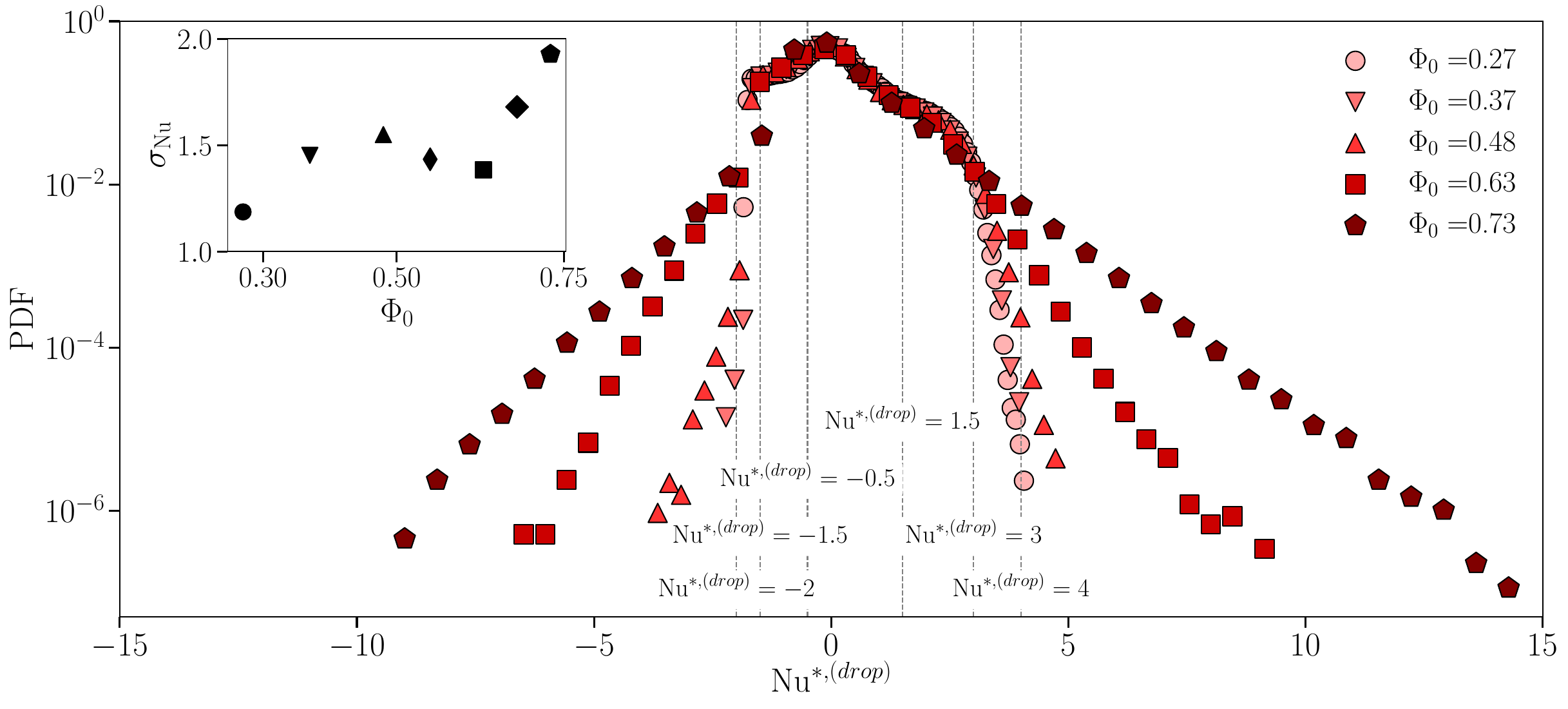}
    \caption{Probability distribution functions (PDFs) of the droplet-scale observable $\mathrm{Nu}_i^{*,(drop)}$, given in Eq.~\eqref{eq:nu_drop_star}. Different symbols/colors correspond to different droplet concentrations $\Phi_0$. Vertical dashed lines identify the thresholds of ranges of $\mathrm{Nu}^{*,(drop)}$ considered in Fig.~\ref{fig:PDF_density}. In the inset we report the values of the standard deviation $\sigma_{\mathrm{Nu}}$ of the droplet Nusselt number $\mathrm{Nu}^{(drop)}$, appearing in Eq.~\eqref{eq:nu_drop_star}, as a function of $\Phi_0$.\label{fig:PDF_nu_drop}}
\end{figure*}
\begin{figure*}[t!]
    \centering
     \includegraphics[width=.75 \textwidth]{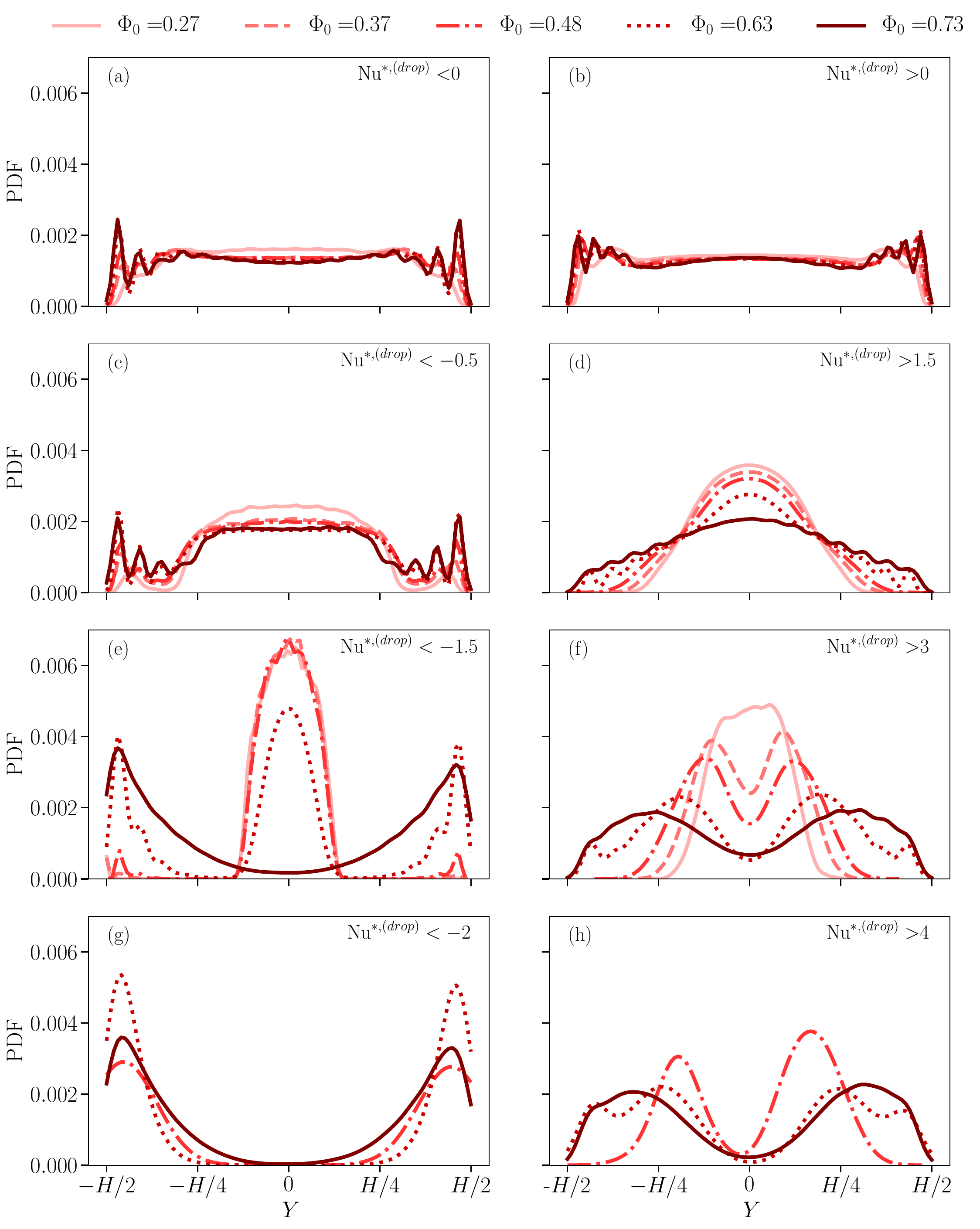}
    \caption{PDF of the center-of-mass position of the droplets along the $y$-direction ($Y$). Different line styles and colors correspond to different droplet concentration $\Phi_0$. Each panel refers to a different range of values of $\mathrm{Nu}^{*,(drop)}$, moving from $\mathrm{Nu}^{*,(drop)} = 0$ towards the extremes negative (left panels) and positive tails (right panels).\label{fig:PDF_density}}
\end{figure*}
\begin{figure*}[t!]
    \centering
    \includegraphics[width=0.75\linewidth]{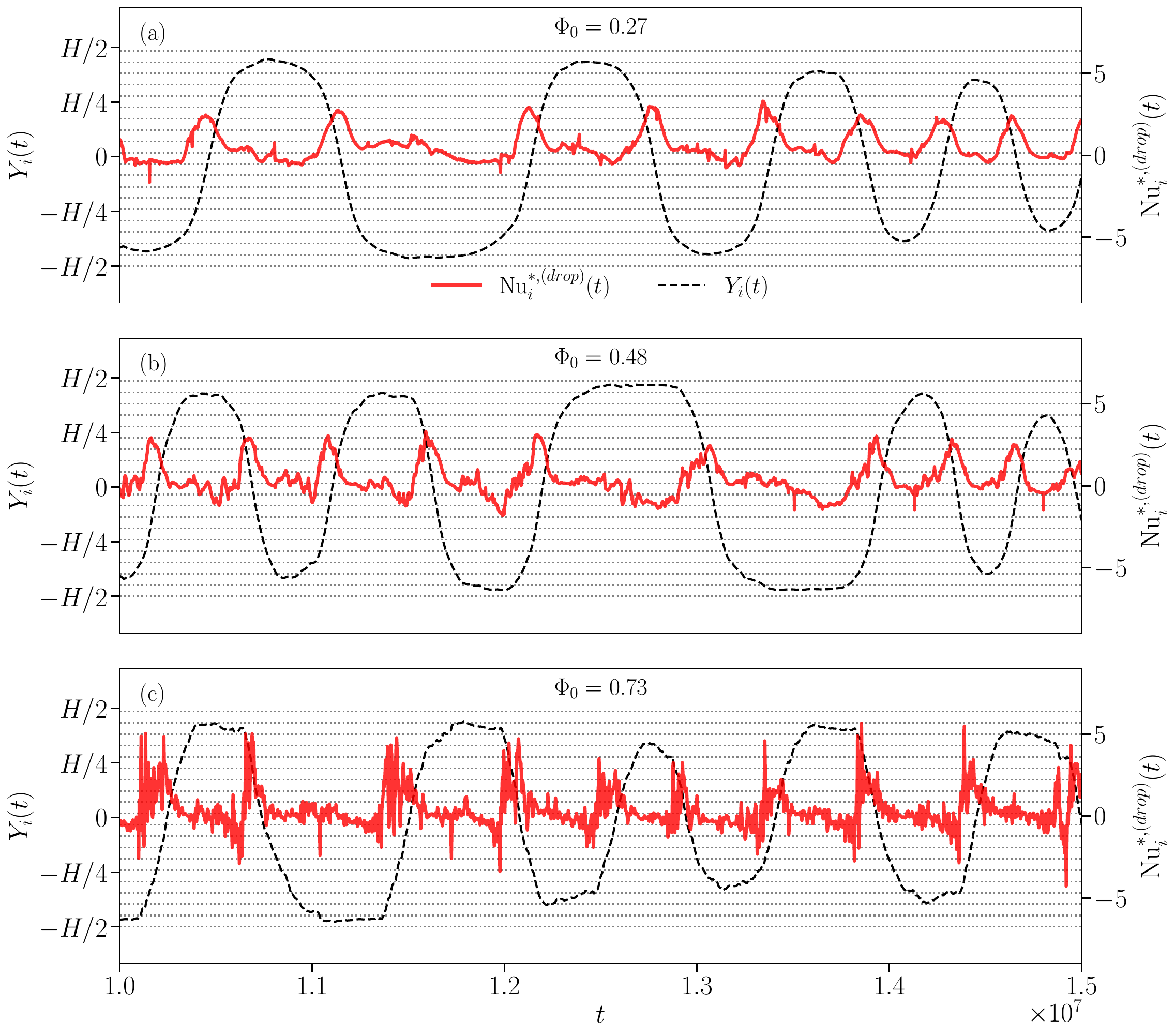}
\caption{Time evolution of $\mathrm{Nu}_i^{*,(drop)}$ (solid red lines) and $Y_i$ (dashed black lines) of a selected droplet $i$, one for each considered droplet concentration $\Phi_0$ (panel (a): $\Phi_0 = 0.27$, panel (b): $\Phi_0 = 0.48$, and panel (c): $\Phi_0 = 0.73$). Time is shown in simulation units. Dotted gray lines outline the droplets layers. In Section~3 of the Supplementary Material the reader can find the corresponding density map animations (Phi027.mp4, Phi048.mp4, and Phi073.mp4).\label{fig:NuYtime}}
\end{figure*}
\section{Results and discussion}\label{sec:analysis}

\begin{figure*}[t!]
    \centering
    \includegraphics[width=0.95 \linewidth]{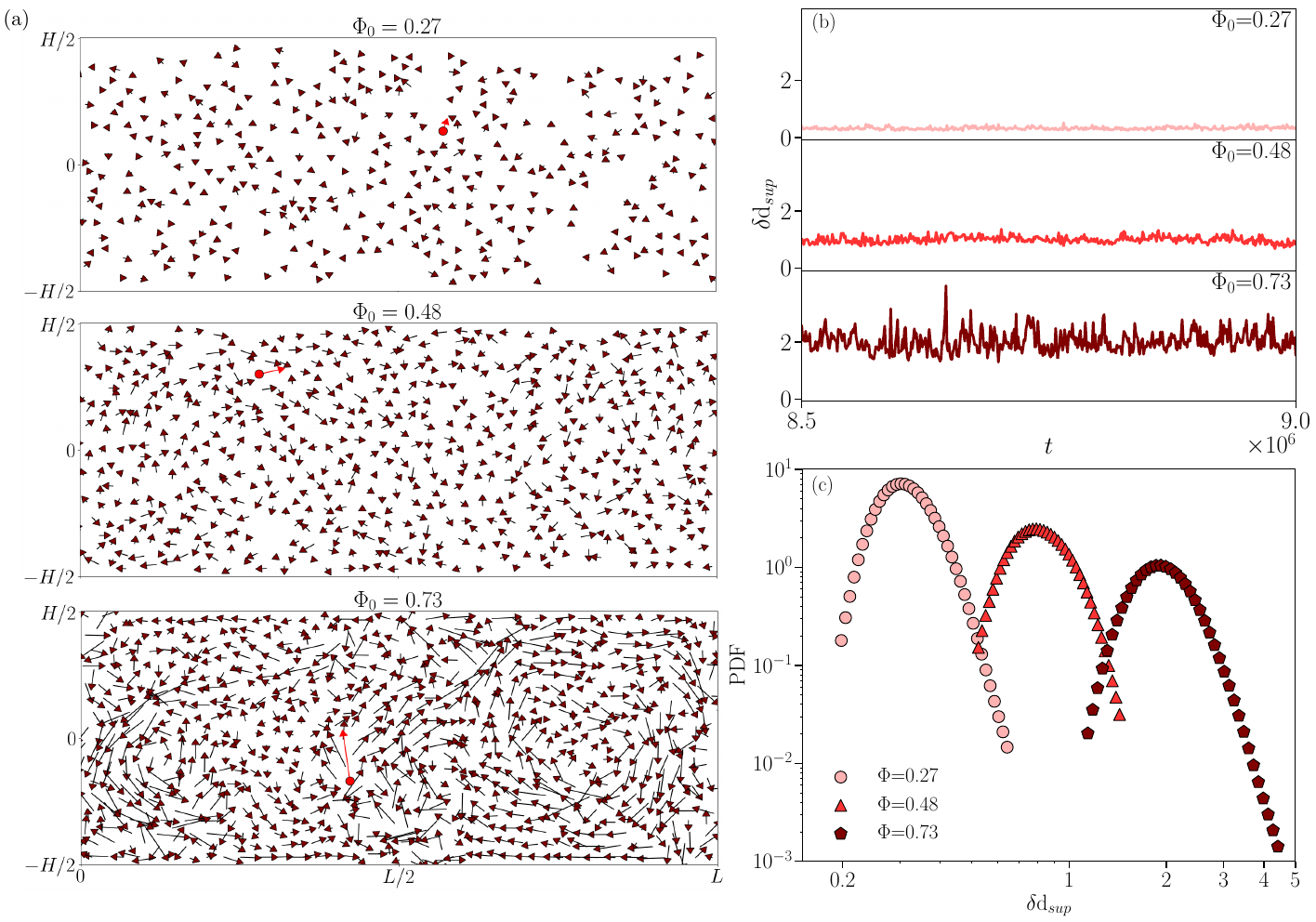}
    \caption{Panel (a): snapshots of the droplet displacement fluctuation $\delta \mbox{d}$ (brown arrows) for different droplet concentrations $\Phi_0$. Red arrows refer to the largest absolute value of $\delta \mbox{d}$ (i.e., the corresponding instantaneous $\delta {\mbox{{\bf d}}}_{sup}$).
    Panel (b): signal in time of the absolute value of $\delta \mbox{{\bf d}}_{sup}$ (i.e., $\delta \mbox{d}_{sup}$) for different $\Phi_0$. In panels (a) and (b), from top to bottom, different sub-panels refer to $\Phi_0 = 0.27$, $\Phi_0 = 0.48$, and $\Phi_0 = 0.73$, respectively. Panel (c): PDFs of $\delta \mbox{d}_{sup}$. Different symbols/colors correspond to different droplet concentration.
    Dimensional quantities are shown in simulation units.\label{fig:d_sup}}
\end{figure*}
We first report on the statistical analysis of the droplet heat flux fluctuations $\mathrm{Nu}_i^{*,(drop)}$ defined in Eq.~\eqref{eq:nu_drop_star}. The aim is to extend and improve the preliminary results reported in Ref.~\cite{PelusiSM21}, which allowed to establish a link between the fluctuations on the macroscopic observable Nu shown in Fig.~\ref{fig:sketch}(b) and the statistics of heat transfer at the droplet scale. In Ref.~\cite{PelusiSM21}, only two ``categories'' of emulsions have been compared, i.e., a  less concentrated Newtonian and a more concentrated non-Newtonian emulsion. Here a systematic characterization at increasing droplet concentration $\Phi_0$ is provided in order to delve deeper into the way heat flux fluctuations emerge, i.e., whether there is a sharp transition or a more continuous change. In Fig.~\ref{fig:PDF_nu_drop}, we show the probability distribution function (PDF) of the droplet-scale observable $\mathrm{Nu}_i^{*,(drop)}$ for different droplet concentrations $\Phi_0$ (different symbols/colors). It is apparent that the larger $\Phi_0$ and the more pronounced are the PDF tails. We also observe that these tails grow continuously upon increasing the droplet concentration. This trend is also confirmed by the inset of Fig.~\ref{fig:PDF_nu_drop}, reporting the values of $\sigma_{\mathrm{Nu}}$ used to normalize the droplet Nusselt number fluctuations (crf. Eq.~\eqref{eq:nu_drop_star}). It shows a non-monotonic behaviour, that may be due to statistical fluctuations, and an almost doubled standard deviation in the more concentrated case as compared to the less concentrated one. Moreover, for a more concentrated emulsion, the authors of Ref.~\cite{PelusiSM21} showed that large fluctuations in the PDF of $\mathrm{Nu}^{*,(drop)}$ are due to droplets localized in the boundary layers. Here we take a step forward in that starting from the PDF reported in Fig.~\ref{fig:PDF_nu_drop} we report the spatial localization of the events that contribute to the PDF tails. Specifically, we compute the PDF of the $y$-coordinate of the droplet center-of-mass positions $Y$, by conditioning the statistics to a specific range of $\mathrm{Nu}^{*,(drop)}$ in Fig.~\ref{fig:PDF_nu_drop}: from the entire positive ($\mathrm{Nu}^{*,(drop)} > 0$) and negative ($\mathrm{Nu}^{*,(drop)} < 0$) tails, to smaller portions of them (e.g., $\mathrm{Nu}^{*,(drop)} > 4$ and $\mathrm{Nu}^{*,(drop)} < -2$). Results are shown in Fig.~\ref{fig:PDF_density}, where an increasing $\Phi_0$ results in different line styles and colors and each panel refers to a different range of $\mathrm{Nu}^{*,(drop)}$. Left panels ((a)-(c)-(e)-(g)) refer to data contributing to the negative tails, whereas right panels ((b)-(d)-(f)-(h)) refer to portions of positive tails. PDFs are close to being flat when considering all collected data (panels (a) and (b)), whereas the scenario changes when approaching the extreme tails. Any emulsion shows a common trend: droplets moving in the proximity of the center of the rolls, do not experience large velocities and follow the average-in-time flow, thus not contributing to large heat transfer fluctuations whereas only droplets which are close to the walls, and receive a boost from the upward or downward thermal plume, instantly experience a significant velocity, resulting in a shift of the peaks of PDF($Y$) from the center of the Rayleigh-Bénard cell towards the boundary layers. This trend is exacerbated in the most concentrated cases, where droplets are highly packed and the free volume of each droplet, as well as their mobility, is, in general, very limited. In other words, the extreme heat transfer fluctuations observed especially in the case of more concentrated emulsions are localized close to the walls of the Rayleigh-Bénard cell. Finally, we can state that Fig.~\ref{fig:PDF_density} is a further confirmation of the continuous transition from less concentrated emulsions, exempted from heat flux fluctuations, to more concentrated emulsions, that accommodate anomalous heat flux fluctuations at the macro- as well as droplet- scales. Note that, in the case of $\Phi_0 = 0.48$, the PDF($Y$) is slightly asymmetric for $\mathrm{Nu}^{*,(drop)}>4$ because of the limited statistics. Moreover, since PDF($\mathrm{Nu}^{*,(drop)}$) in Fig.~\ref{fig:PDF_nu_drop} are asymmetric, then the shape of PDF($Y$) for the case $\mathrm{Nu}^{*,(drop)}< -2$ is qualitatively similar to the one for $\mathrm{Nu}^{*,(drop)}> 5$ (data not shown). To enrich this result with information about the dynamics, we monitored the time evolution of $\mathrm{Nu}_i^{*,(drop)}$ and the corresponding $Y_i$ for a single droplet $i$ randomly chosen among droplets moving closer to the walls (see Fig.~\ref{fig:NuYtime}). For the sake of readability, we show data for only three values of $\Phi_0$ (panel (a): $\Phi_0 = 0.27$, panel (b): $\Phi_0 = 0.48$, and panel (c): $\Phi_0 = 0.73$). From Fig.~\ref{fig:NuYtime} it is clear that $\mathrm{Nu}_i^{*,(drop)}$ exhibits an intermittent behavior, more pronounced as the droplet concentration increases. It reveals a non-trivial correlation between ``bursts" in the droplet heat transfer fluctuation and the spatial approach-to/departure-from a wall. Contrariwise, when the droplet ``slips" close to the wall, i.e., in the periods where the signal of $Y_i$ is almost flat, the droplet does not contribute to an extreme event. In addition, Fig.~\ref{fig:NuYtime} reveals that the oscillation period of $Y_i$ is dependent on the droplet concentration and it is not constant as expected for a homogeneous material. We interpret this finding as due to a droplet layer change by the drop under consideration, whose oscillation period around the center of the convective roll decreases as the droplet approaches it (see dotted lines outlining droplets layers). The droplet layer change is triggered by one or more collisions with the surrounding droplets, especially in the concentrated case where the droplet mobility is further reduced. This picture can be caught by the eye by watching the density map animations we include in Section~3 of the Supplementary Material (Phi027.mp4, Phi048.mp4, and Phi073.mp4). Notice that in the less concentrated case the selected droplet rarely moves in the boundary layers, as already observed in Fig.~\ref{fig:PDF_density}.\\

\begin{figure*}[t!]
    \centering
    \begin{tabular}{c c}
    \includegraphics[width=0.43\linewidth]{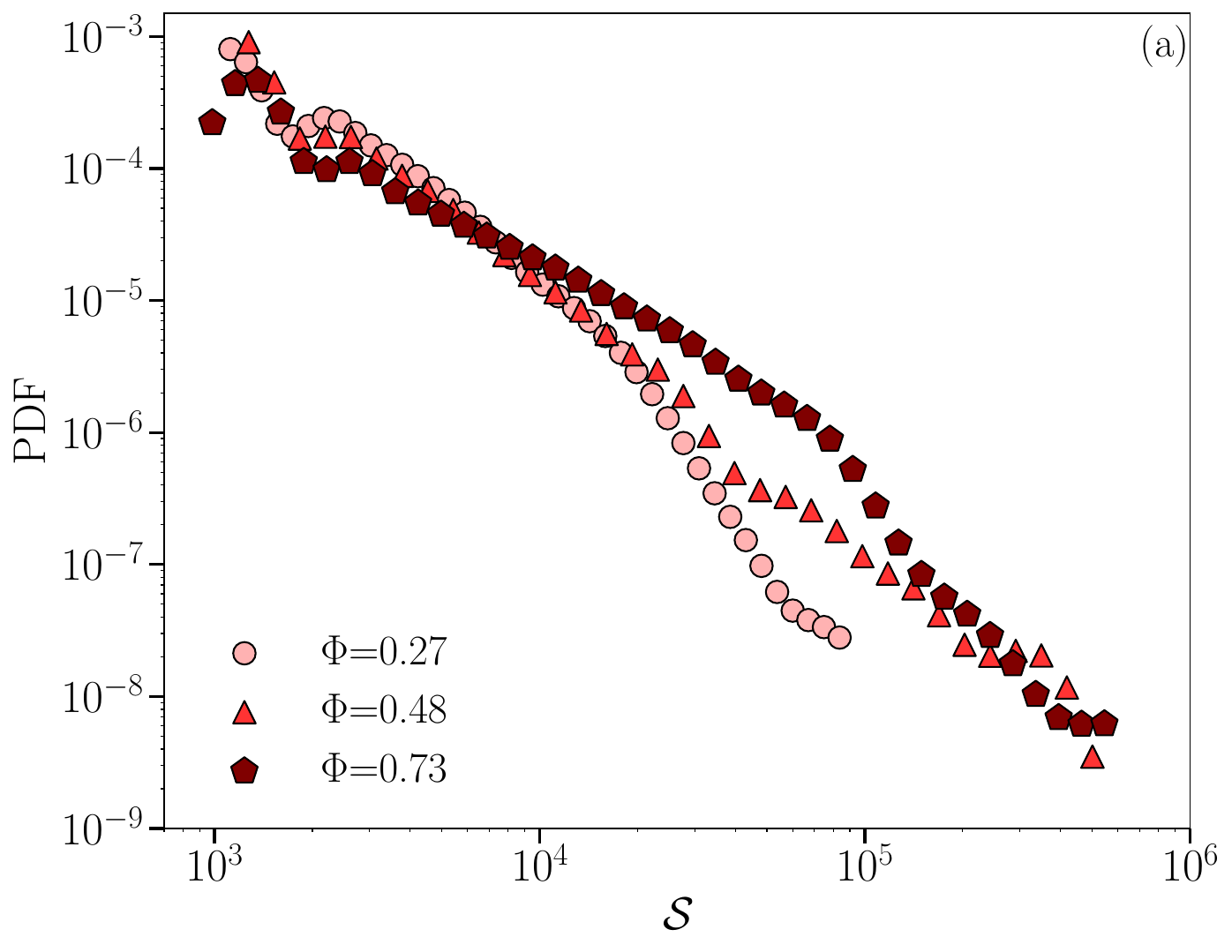} & \includegraphics[width=0.43\linewidth]{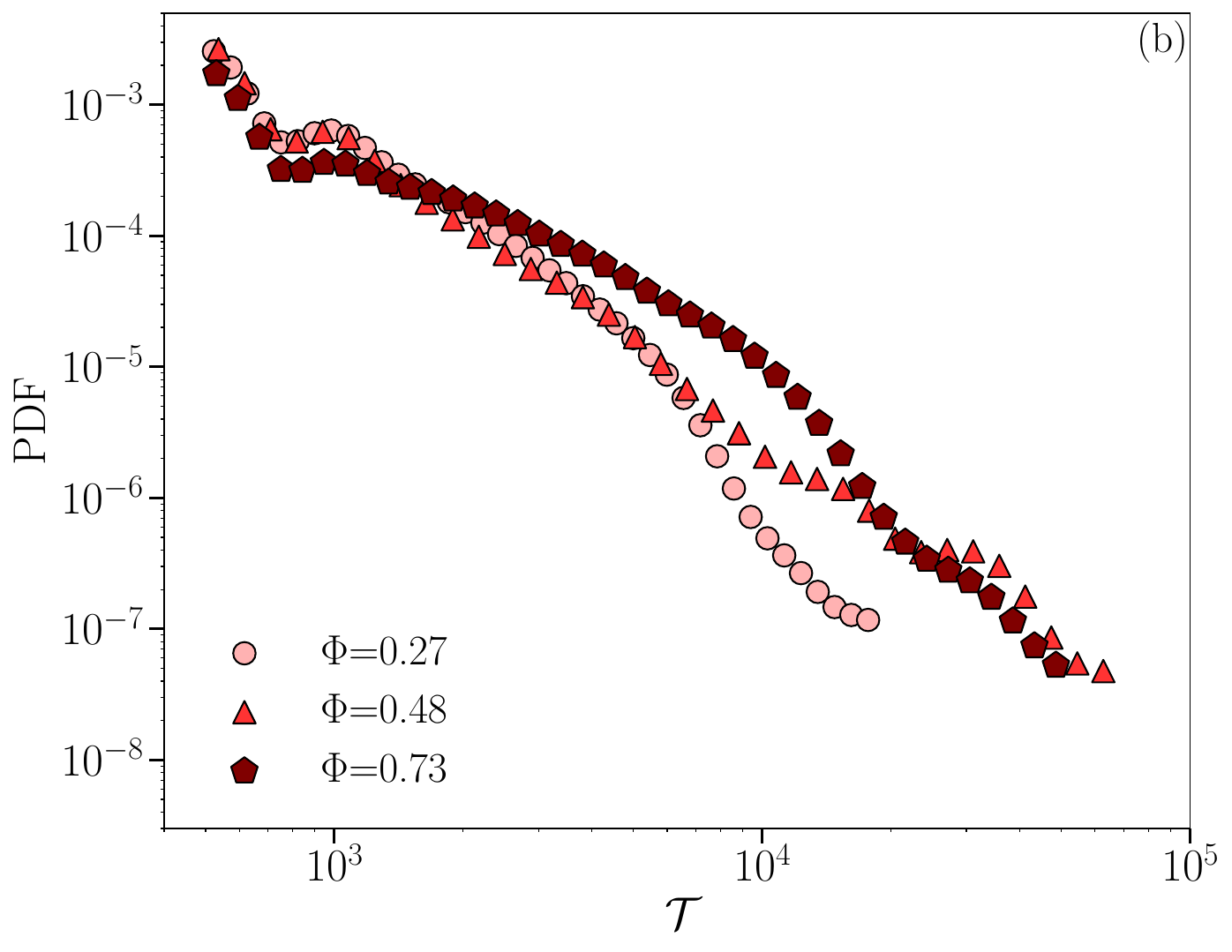}
    \end{tabular}
    \caption{Statistics of the properties of coherent droplets motion. Panel (a): PDFs of the spatial extension $S$. Panel (b): PDFs of the temporal duration $\mathcal{T}$. Different symbols/colors correspond to different droplet concentrations $\Phi_0$. $S$ and $\mathcal{T}$ are shown in simulation units.\label{fig:avalanches}}
\end{figure*}

\begin{figure*}[t!]
    \centering
    \begin{tabular}{c c}
    \includegraphics[width=0.43\linewidth]{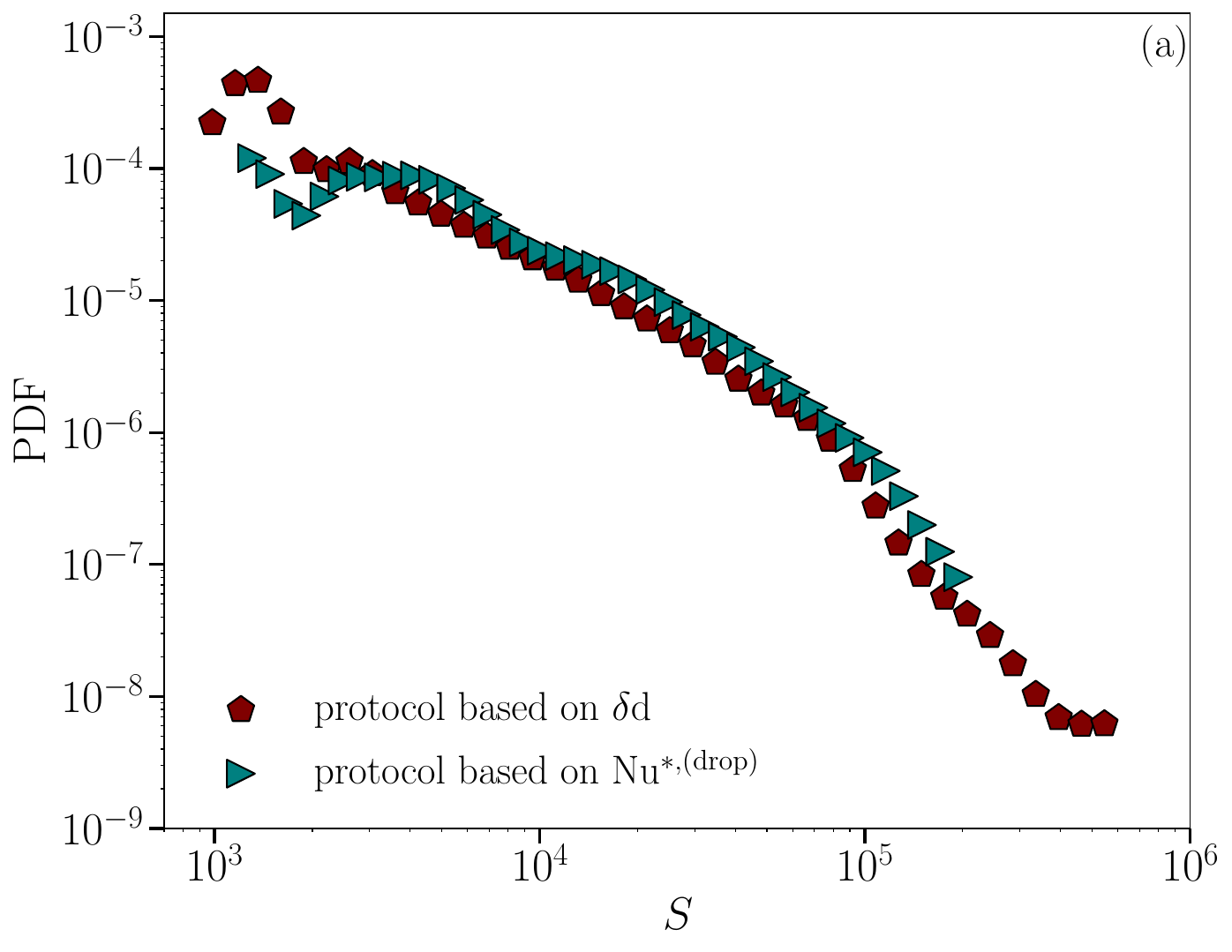} & 
    \includegraphics[width=0.43\linewidth]{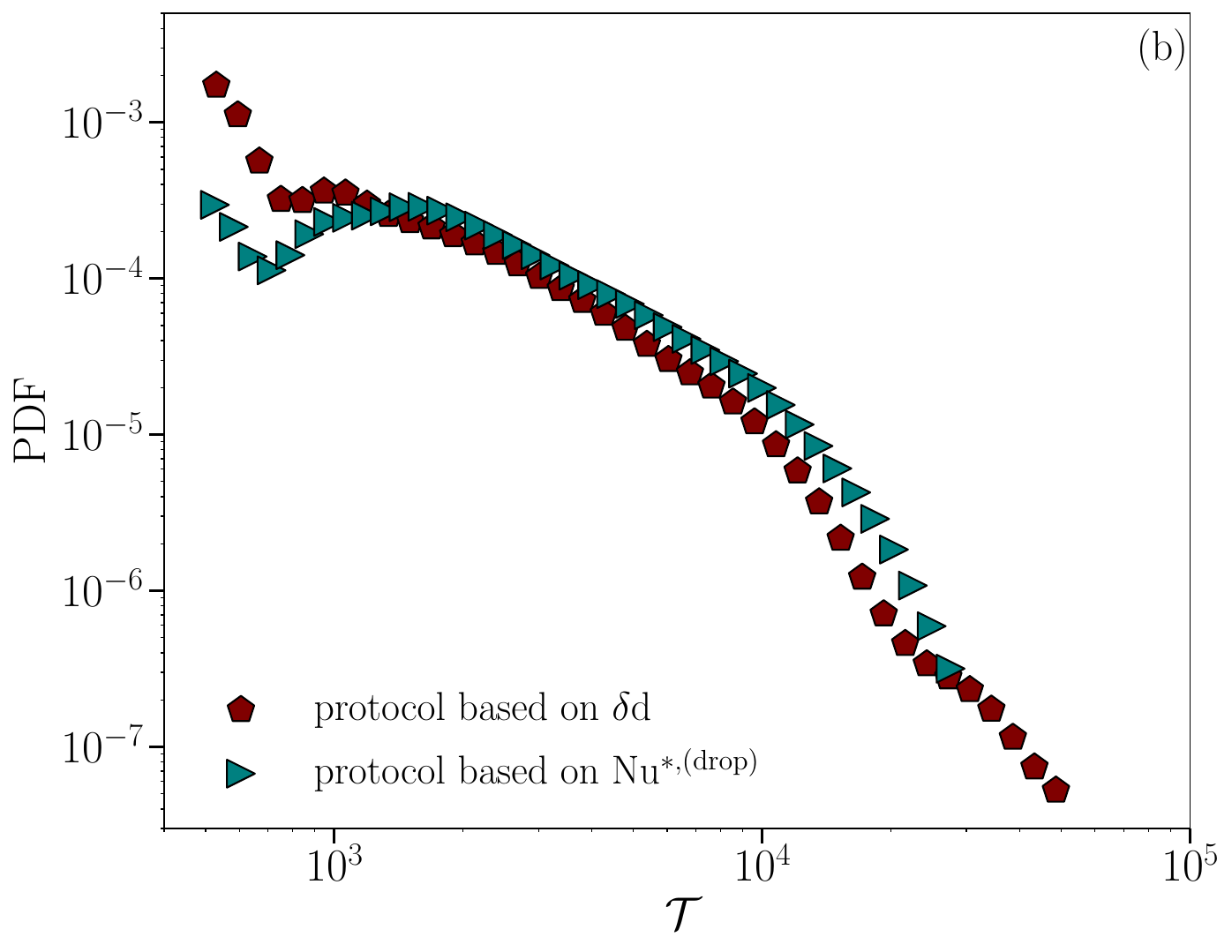}
    \end{tabular}
    \caption{A comparison between the PDF for the spatial extension $S$ of the coherent droplets motion (panel (a)) and its duration $\mathcal{T}$ (panel (b)) computed by using a protocol based on $\delta {\bf d}$ (brown pentagons) and $\mathrm{Nu}^{*,(drop)}$ (teal rightward-pointed triangles) for the most packed emulsion ($\Phi_0 = 0.73$). $S$ and $\mathcal{T}$ are shown in simulation units.\label{fig:avalanches_protocol}}
\end{figure*}

The observation of an intermittent behavior in the heat transfer and its fluctuations in RBC of concentrated emulsions naturally leads to asking whether these materials present a droplets collective motion in such a situation. To answer this question, we inspected the statistical properties of droplets coherent motion in terms of its spatial extension $S$ and its temporal duration $\mathcal{T}$ following Ref.~\cite{PelusiAvalanche19}. In the latter reference, the system was driven by coarsening dynamics, characterized by an averaged-in-time flow that is almost zero, and the authors employed a protocol based on the absolute value of the vectorial droplet displacement ${\bf d}_i(t)$ (cfr. Fig.~\ref{fig:sketch}(c), black arrows). However, under RBC the system's dynamics exhibits an averaged-in-time flow characterized by the presence of convective rolls (cfr. Fig.~\ref{fig:sketch}(c), colorbar), meaning that, to apply the same protocol, we need to subtract the mean flow and consider the vectorial droplet displacement fluctuation $\delta {\bf d}$ as the key observable (cfr. Eq.~\eqref{eq:displacementFluct} and Fig.~\ref{fig:sketch}(c), brown arrows). As highlighted by Fig.~\ref{fig:d_sup}(a), this observable exhibits a weak intensity in the less concentrated emulsion with no manifest coherence, while it is substantial in the more concentrated case, highlighting large regions of droplet coherent motion. Thus, in order to focus on extreme events, we select at any time step only the droplet with the largest absolute value of $\delta {\bf d}$, obtaining an intermittent signal in time of $\delta {\mbox{d}}_{sup}$ (cfr. Fig.~\ref{fig:d_sup}(b)). As threshold value of $\delta {\mbox d}$ to identify a droplets coherent motion, we take the knee value of the PDF of that observable (cfr. Fig.~\ref{fig:d_sup}(c)): whenever the signal of $\delta {\mbox{d}}_{sup}$ overcomes the threshold, we record a $t_{start}$ as the beginning of the spatial coherence. Then, the first time $t>t_{start}$ where $\delta {\mbox{d}}_{sup}$ returns below the threshold, we record its end ($t_{end}$). We measure the spatial extension of the detected coherent motion $S$ by summing up the area of droplets whose absolute value of $\delta {\bf d}$ is larger than the threshold value during its temporal duration, defined as $\mathcal{T} =t_{ end} - t_{ start}$. The statistics of $S$ at varying droplet concentration $\Phi_0$ is reported in Fig.~\ref{fig:avalanches}(a): it is evident how the extension of the PDF of $S$ increases as droplet concentration $\Phi_0$ increases, meaning that the spatial coherence develops in originally small regions which grow in extension with $\Phi_0$. However, only the most concentrated emulsion presents a robust power-law behavior that covers two decades, echoing observations on other systems presenting an avalanche-like behaviour~\cite{Barkhausen17,Ruina83,BonamySantucciPonson08}. This result is further confirmed by analyzing the statistics of the temporal duration $\mathcal{T}$ (cfr. Fig.~\ref{fig:avalanches}(b)), where, once again, a power-law behavior is better visible in the case of the most concentrated emulsion. Both panels of Fig.~\ref{fig:avalanches} still confirm how the phenomenology changes in a continuous way from less to more concentrated emulsions. We remark that the PDF cut-off at large values of $S$ and $T$ arise because of system size, in agreement with Ref~\cite{Leishangthem17,PelusiAvalanche19}. In addition, since $\delta {\bf d}$ is not the only observable at the droplet-scale we can measure in our simulations, it is interesting to double-check whether the droplet-scale heat transfer fluctuation $\mathrm{Nu}^{*,(drop)}$ is equally good at capturing the droplets coherent motion of emulsions under RBC. Fig.~\ref{fig:avalanches_protocol} provides the answer, showing the comparison of the PDF of spatial coherence $S$ and duration $\mathcal{T}$ for $\Phi_0 = 0.73$ between the protocol based on $\delta {\bf d}$ and the one based on $\mathrm{Nu}_i^{*,(drop)}$, where the latter follows the same procedure mentioned for $\delta {\bf d}$. The qualitative good overlap of the two PDFs is somehow expected but it contains multiple information: first, it confirms the strong correlation between bursts of heat transfer and displacement at the droplet scale; second, it also supports the idea of the emergence of ``thermal avalanches'' for the most packed systems and very close to the transition point between conduction and convection; then, it confirms that our analysis and simulations are realistic in reproducing the same qualitative results at changing the observable at the droplet scale.

\section{Conclusions}\label{sec:conclusions}

We have used numerical simulations with the open source code TLBfind~\cite{TLBfind22} to study the statistical properties of heat flux fluctuations in concentrated emulsions under thermal convection, just above the transition from conductive to convective states. Numerical simulations have been helpful in providing a systematic analysis at varying droplet concentrations, ranging from less concentrated (showing small heat transfer fluctuations) to more concentrated emulsions (highlighting enhanced heat flux fluctuations). By systematically increasing the droplet concentration, we have observed evidence of a continuous transition that goes together with a continuous rise of the tails of the PDF of the droplet-scale observable $\mathrm{Nu}^{*,(drop)}$ used to quantify the droplet heat transfer fluctuations. We have analyzed these extreme fluctuations, observing their strong correlation with the droplet spatial localization, resulting in an accumulation in the boundary layer region of droplets contributing to the extreme fluctuations. The physical effect hinges on the presence of finite-size droplets, which increase their contact as the droplet concentration increases but without merging via coalescence events. These ``soft collisions'' provoke the observed extreme heat transfer fluctuations and collective phenomena, such as the coherent droplets motion with a spatial extension $S$ and with a characteristic temporal duration $\mathcal{T}$. A statistical analysis of both $S$ and $\mathcal{T}$ has been conducted, showing a neat power-law behavior when the emulsion is more concentrated. This result is found to be robust at changing the droplet-scale observable used to identify a droplet coherent motion.\\
We remark that the observed findings of anomalous fluctuations have the peculiarity of emerging while the emulsion experiences a laminar regime. This is unexpected in a homogeneous fluid dynamics, since they are observed appearing along with the onset of a turbulent regime~\cite{Welti05,Shang05,Scheel14,Zonta16}. It is apparent that a more precise analysis would be needed for a real quantitative assessment. This prompts the need for a more clear comparison between the two dynamics (laminar emulsion vs. turbulent homogeneous fluid), which surely sets interesting future perspectives. Our findings also open new questions concerning other aspects that may affect the heat transfer properties in emulsion systems with finite-size droplets. For instance, it is not known so far if the heat flux fluctuations explored in this work can be more pronounced by changing from a two-dimensional to a three-dimensional system or by varying the number of convective rolls in the Rayleigh-Bénard cell. The role played by these fluctuations may also change as the Rayleigh number $Ra = \rho \alpha g \Delta T H^3/(\kappa \eta_0)$ increases, assessing the regime that for Newtonian fluids is no longer laminar. Furthermore, an investigation into the effect of confinement may be useful to shed light on the characteristic domain size limit beyond which the observed extreme heat flux fluctuations disappear.

\section*{Acknowledgements}

We wish to acknowledge Fabio Bonaccorso for his support. MS and MB gratefully acknowledge CN1 -- Centro Nazionale di Ricerca in High-Performance Computing Big Data and Quantum Computing -- for support. This work received funding from the European Research Council (ERC) under the European Union’s Horizon 2020 research and innovation program (grant agreement No 882340).

\bibliographystyle{rsc}
\bibliography{francesca}

\end{document}